\documentclass[arxiv]{aa}

\usepackage{ulem}
\usepackage{graphicx}
\usepackage{txfonts}
\usepackage{amsmath}
\usepackage{amssymb}
\usepackage{subcaption}
\usepackage{multirow}
\usepackage{adjustbox}
\usepackage{comment}
\usepackage{tabto}

\usepackage[colorlinks,breaklinks]{hyperref}
\hypersetup{linkcolor=blue,citecolor=blue,filecolor=black,urlcolor=blue}

\usepackage[switch]{lineno}
\nolinenumbers

\newcommand{\nickel}{$^{56}$Ni}
\newcommand{\prometal}{$Z_{progenitor}$}
\newcommand{\solarmetal}{$Z_{\odot}$}
\newcommand{\solarmass}{$M_{\odot}$}
 \newcommand{\chmass}{$M_{ch}$}

\newcommand{\proafe}{\textit{$(\alpha/Fe)_{progenitor}$}}

\newcommand{\orcid}[1]{\href{https://orcid.org/#1}{\includegraphics[height=11pt]{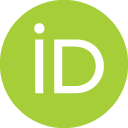}}}

\defcitealias{Kim2024}{K24}
\defcitealias{Timmes2003}{TBT03}
\defcitealias{Howell2009}{H09}
\defcitealias{Neill2009}{N09}
\defcitealias{Bravo2010}{B10}
\defcitealias{Leung2018}{LN18}
\defcitealias{Leung2020}{LN20}
\defcitealias{Gronow2021}{G21}

\begin{document} 

\title{Testing the effect of progenitor's metallicity on \nickel{} mass and constraining the progenitor scenarios in Type Ia supernovae}

\titlerunning{Constraining SN Ia Progenitor Scenarios on the \prometal{}--\nickel{} Mass Diagram}
\authorrunning{Y.-L. Kim et al.}

\author{
	Young-Lo~Kim\inst{1, 2}\thanks{ylkim83@yonsei.ac.kr}\orcid{0000-0002-1031-0796},
	Chul Chung\inst{1, 2}\orcid{0000-0001-6812-4542},
	Yong -Cheol Kim$^{1}$\orcid{0000-0002-1907-0848}
}

\institute{
	Department of Astronomy, Yonsei University, Seoul 03722, Republic of Korea
	\and 
	Center for Galaxy Evolution Research, Yonsei University, Seoul 03722, Republic of Korea
	}

\date{Received April 01, 2025; accepted xxx xx, xxxx}
 
\abstract {
The analytical model found that the intrinsic variation in the initial metallicity of the Type Ia supernova (SN Ia) progenitor stars (\prometal{}) translates into a 25\% variation in the \nickel{} mass synthesized and, therefore, $\sim$0.2 mag difference in the observed peak luminosity of SNe Ia.
Previous observational studies used the currently-observed global gas-phase metallicity of host galaxies, instead of \prometal{} used in the model, and showed a higher scatter in the \nickel{} mass measurements compared to the model prediction. 
Here, we use \prometal{} of 34 normal SNe Ia and employ recent SN Ia explosion models with various configurations to cover the observed \nickel{} mass range. 
Unlike previous studies, which only have samples in the sub-solar range, our sample covers the \prometal{} range ($\frac{1}{3}$ $Z_{\odot}$ < \prometal{} < 3 $Z_{\odot}$), where most of the \prometal{} effect occurs. 
Linear regression returns a slope of $0.02\pm0.03$, which is the opposite trend to the analytical model, but at at low statistical significance level. 
We find that comparing our sample with SN Ia explosion models on the \prometal{}--\nickel{} mass diagram allows us to constrain the progenitor scenarios. 
We also explore other chemical composition indicators, such as $(Fe/H)_{progenitor}$ and $(\alpha/Fe)_{progenitor}$.
For $(Fe/H)_{progenitor}$, our sample follows the trend predicted by the analytical models, but at a low significance level (0.4$\sigma$). 
Noticeably, $(\alpha/Fe)_{progenitor}$ shows the opposite trend and a clear gap. 
When we split the sample at $(\alpha/Fe)_{progenitor}$ = 0.35 $(\alpha/Fe)_{\odot}$, we find a 3$\sigma$ difference in the weighted-means of the \nickel{} mass.
Lastly, SNe Ia in different \prometal{} groups show a difference of $0.14\pm0.09$ ($1.6\sigma$) mag in the standardized luminosity. 
The present work highlights a holistic approach (from the progenitor star to the explosion with SN Ia and host galaxy observational data) to understand the underlying physics of SNe Ia for more accurate and precise cosmology.
}


\maketitle

\section{Introduction}
\label{sec:intro}

The peak luminosity of Type Ia supernovae (SNe Ia) is an essential ingredient for their use as distance indicators.
The empirical relationship between the peak luminosity and its decline rate \citep{Phillips1993} and also its peak colour \citep{Tripp1998} reduced the intrinsic variation in the SN Ia peak luminosity in the $V$-band from $\sim$1 mag  to $\sim$0.1 mag.
From this accuracy, SNe Ia provided direct observational evidence for the accelerating expansion of the universe \citep{Riess1998, Perlmutter1999}.
Recently, combined with other cosmological probes, it was found some challenges in cosmology, such as the "Hubble Tension" \citep{Riess2022} and preference for dynamical dark energy \citep{DESI2025}.
Although SNe Ia play an important role in cosmology, the physical origin of the empirical relationship and the peak luminosity variation in SNe Ia remains elucidate.
Therefore, understanding this would help alleviate the challenges we are currently facing.

The intrinsic variation in the SN Ia peak luminosity, theoretically, is tightly correlated with the amount of \nickel{} synthesized during the explosion, because the explosion of SNe Ia is powered by the radioactive decay of \nickel{} \citep{Truran1967, Colgate1969}.
This \nickel{} amount presumably depends on the properties of the progenitor star (e.g., mass and metallicity) and on the details of the explosion scenarios (near-Chandrasekhar/sub-Chandrasekhar mass ($M_{ch}$) and deflagration/detonation explosions).
In this regard, \citet[][hereafter \citetalias{Timmes2003}]{Timmes2003} analytically explored how the intrinsic variation in the initial metallicity of the SN Ia progenitor stars (\prometal{}) translates into a variation in the \nickel{} mass synthesized and, therefore, the peak luminosity of SNe Ia.

\citetalias{Timmes2003} assumed that nearly all one-dimensional $M_{ch}$ models of SNe Ia produce most of the \nickel{} in a burn to nuclear statistical equilibrium between the mass shells of 0.2 and 0.8 \solarmass{}.
They found the linear relationship between \prometal{} and the \nickel{} mass:

\begin{equation}
\label{eq:tbt03}
$\nickel{}$ \text{ mass} \approx  0.6 M_{\odot} \left( 1 - 0.057 \dfrac{Z_{progenitor}}{Z_{\odot}} \right).
\end{equation}

This is because higher-metallicity main-sequence stars expect to evolve into white dwarfs (WDs) with more neutron-rich elements, producing more stable burning products relative to radioactive \nickel{}.
\citetalias{Timmes2003} suggested that the progenitor metallicity could account for a 25\% variation in \nickel{} mass synthesized in SNe Ia and thus $\sim$0.2 mag in the observed peak luminosity of SNe Ia in the $V$-band.
 
After \citetalias{Timmes2003}, several studies have been conducted to update this analytical model.
\citet[][hereafter \citetalias{Howell2009}]{Howell2009} accounted for the fact that $O/Fe$ can vary as a function of $Fe/H$, such that [O/Fe] = $a$ + $b$ [Fe/H] taken from \citet{Ramirez2007}, while \citetalias{Timmes2003} assumed that $O/Fe$ is constant relative to $Fe/H$.
\citet{Ramirez2007} fitted this linear relation and provided coefficients of $a$ and $b$ for different populations of stars in the thin disk, the thick disk, and the halo (see their table 3 for $a$ and $b$ values of each population).
\citetalias{Howell2009} considered those different populations of stars when calculating $O/Fe$ via different $a$ and $b$ values.
Then, with [O/Fe] = [O/H] $-$ [Fe/H], they suggested a linear relation between the \nickel{} mass synthesized in SNe Ia and $O/H$ and also $Fe/H$\footnote{Here, we are using the standard notation: $[A/B] \equiv \log_{10}(A/B) - \log_{10}(A/B)_\odot$.}:

\begin{multline}
\label{eq:h09_ofe}
\dfrac{M_{56}}{M^{0}_{56}} = 1 - 0.044 \left[ \frac{(O/H)}{10^{-3}} \right]  \left\{ 1 + 0.122 \left[ \frac{(O/H)}{10^{-3}} \right]  \right. \\  
\left. + 10^{-(0.19+0.53b+a)/(1+b)} \left[ \frac{(O/H)}{10^{-3}} \right] ^{-b/(1+b)} \right\},
\end{multline}

\begin{multline}
\label{eq:h09_feh}
\dfrac{M_{56}}{M^{0}_{56}} = 1 - 0.020 \times 10^{a+(1+b)[Fe/H]} \{1 + 0.056 \times 10^{a+(1+b)[Fe/H]}  \\
+ 0.64 \times 10^{-a-b[Fe/H]} \}.
\end{multline}

Here, $M^{0}_{56}$ is the \nickel{} mass synthesized at an electron abundance equalls to $\frac{1}{2}$, i.e., for a pure C--O WD. 
We note that \citetalias{Howell2009} used the \citet{Asplund2005} abundances for the solar composition.

Another update of \citetalias{Timmes2003} is by \citet[][hereafter \citetalias{Bravo2010}]{Bravo2010}. They suggested a steeper slope of 0.075 (see their equation 2) than that of \citetalias{Timmes2003} based on a series of SNe Ia explosion simulations, considering the $M_{ch}$ deflagration-to-detonation transition models with a central density of a WD ($\rho_{c}$) = $3 \times 10^{9}$ g cm$^{-3}$ and \prometal{} = 10$^{-5}$ -- 0.10 \solarmetal{}.
They also explored a non-linear relation considering the deflagration-to-detonation transition density as a function of the local \prometal{}:

\begin{equation}
\label{eq:bravo10non}
$\nickel{}$ \text{ mass} \propto 1 - 0.18 \dfrac{Z_{progenitor}}{Z_{\odot}} \left( 1 - 0.10 \dfrac{Z_{progenitor}}{Z_{\odot}} \right).
\end{equation}

To test the \citetalias{Timmes2003} model with observational data, there are some studies, for example, \citet{Gallagher2005};  \citetalias{Howell2009}; \citet{Neill2009, MorenoRaya2016}.
Those studies used the average metallicity of the SN Ia host galaxies or the regions around the SN Ia explosion site rather than \prometal{}, because \prometal{} is difficult to determine through observation.
\citet{Gallagher2005} used spectroscopic line ratios of 30 host galaxies to determine metallicity and found that the correlation was at the 70\% confidence level.
\citetalias{Howell2009} and \citet{Neill2009} tested the model with more samples at intermediate- and low-redshift ranges, respectively.
They inferred the metallicity from the stellar mass of host galaxies using the mass-metallicity relation of \citet{Tremonti2004} and \citet{Lee2006}, and concluded that their data were consistent with the \citetalias{Timmes2003} prediction updated by \citetalias{Howell2009}.
\citet{MorenoRaya2016} observed and estimated the metallicity of the 28 regions where the SNe Ia exploded.
Instead of using \nickel{} mass, they used the absolute magnitude of SNe Ia, which is positively correlated to the \nickel{} mass.
They found an 80\% chance of the correlation between the SN Ia absolute magnitude and the local metallicity, in the sense that SN Ia luminosities tend to be higher for galaxies with lower metallicities.
All previous studies using observational data of host galaxies or regions around SN Ia explosion sites presented the same trend as the \citetalias{Timmes2003} model, despite the higher scatter in \nickel{} mass measurements than the \citetalias{Timmes2003} prediction.

More precisely, previous studies used the currently-observed global gas-phase metallicity (e.g., $12 + log(O/H)$).
A recent study by \citet[][hereafter \citetalias{Kim2024}]{Kim2024} showed that there is a difference in metallicity between the SN Ia birth environment and the currently-observed status of host galaxies because of the delay time between the birth of the progenitor star and the SN Ia explosion.
This means that the currently-observed global metallicity used in previous studies would not be representative of \prometal{} (see also a discussion in section 5 of \citealt{Howell2009}).
In addition, \citetalias{Timmes2003} defined metallicity in their paper as CNO + Fe abundances, like total stellar metallicity.
Although the total stellar and gas-phase metallicity are correlated \citep[e.g.,][]{CidFernandes2005}, they are not identical.
Lastly, previous studies do not have enough samples at the higher-metallicity range, where most progenitor metallicity effect occurred.

Furthermore, regarding the higher scatter in \nickel{} mass measurements than predicted by the \citetalias{Timmes2003} model, this implies that the analytical model cannot fully account for the observed \nickel{} production of SNe Ia.
However, recent hydrodynamic simulations of SN Ia explosion models with different explosion mechanisms and progenitor properties, including the \prometal{} dependence, can produce the observed \nickel{} mass range \citep[e.g., ][]{Leung2018, Leung2020, Gronow2021}.
In addition to the \nickel{} abundance, those studies provide most of all the nucleosynthesis yields of SNe Ia in different explosion scenarios. 
Combining the SN Ia yields with the galactic chemical evolution model allows the prediction of the evolution of elemental abundances in the Milky Way and dwarf spheroidal galaxies. 
Then, this prediction is compared with observational data to constrain the SN Ia progenitors \citep[e.g.,][]{Kobayashi2020, Palla2021, Trueman2025}.

In this work, different from previous studies, we will use the total metallicity of the birth environments of the SN Ia progenitor stars determined by \citetalias{Kim2024}.
They, for the first time, determined [Fe/H] and [$\alpha$/Fe] of birth environments through empirical correlations between the stellar population age and [Fe/H] and [$\alpha$/Fe] of galaxies and accurately determined stellar population properties of genuine early-type host galaxies in the redshift range of 0.01 < z < 0.08. 
A genuine early-type galaxy is homogeneous in terms of the stellar population because its stellar population is formed through a single burst of star formation and then followed by passive evolution \citep[e.g.,][]{Thomas2005}.
If an SN Ia is observed in this galaxy, it is most likely that the SN Ia progenitor star formed simultaneously with the formation of the single stellar population in the galaxy.
Therefore, [Fe/H] and [$\alpha$/Fe] of the birth environments determined by \citetalias{Kim2024} can be considered those of the progenitor stars.
These properties would be more representative of the progenitor star than the currently-observed gas-phase metallicity used in previous studies.

With this progenitor star data, we will test the trend of the \nickel{} mass synthesized during the SN Ia explosion as a function of \prometal{} predicted by the analytical model of \citetalias{Timmes2003} and its updated models by \citetalias{Howell2009} and \citetalias{Bravo2010}.
One step further, we will attempt to constrain the SN Ia progenitors (of each SN Ia) by comparing our data with recent SN Ia explosion simulations from \citet{Leung2018}, \citet{Leung2020}, and \citet{Gronow2021}

In Sec.~\ref{sec:method}, we describe our sample and the method to estimate \prometal{} from birth environments of progenitor stars and the \nickel{} mass from the SN Ia light-curve data.
Also, we list and explain recent hydrodynamic simulations of SN Ia explosion models used in the present work.
Then, in Sec.~\ref{sec:result}, we test the relation between \prometal{} and the \nickel{} mass and constrain SN Ia progenitors for our sample on the \prometal{}--\nickel{} mass diagram.
We discuss our result and offer our concluding remarks in Sec.~\ref{sec:discussion}.

\section{Methodology}
\label{sec:method}

\subsection{Sample}
\label{subsec:sample}

\citetalias{Kim2024} determined the SN Ia progenitor star's birth environments, specifically [Fe/H] and [$\alpha$/Fe].
For this, they employed empirical correlations between the stellar population age and [Fe/H] and [$\alpha$/Fe] of galaxies, and accurately determined stellar population properties of SN Ia host galaxies, such as age and metallicity.
The empirical correlations are taken from \citet{Walcher2015}, who provided them based on an analysis of 2286 early-type galaxies. 
Host galaxy properties are taken from \citet{Kang2016, Kang2020}.
They observed 51 nearby (0.01 < $z$ < 0.08) early-type host galaxies from the SN Ia catalogue collected by \citet{Kim2019} and obtained very high-quality spectra (mean signal-to-noise ratio $\sim$175).
Based on absorption line analysis using the Yonsei Evolutionary Population Synthesis model \citep[hereafter YEPS]{Chung2013}, host galaxy stellar population age and metallicity are accurately determined.
Among 51 early-type host galaxies, \citetalias{Kim2024} only used 44 genuine early-type galaxies, which were identified from an UV-optical-IR colour-colour diagram by \citet{Kang2020}.

Out of the 44 host galaxies in the \citetalias{Kim2024} sample, we used 34 hosts of cosmologically normal SNe Ia.
They are selected based on the typical cut criteria when making a cosmological sample, such as SALT2  \citep{Guy2007, Guy2010} $|x_{1}| < 3$, $|c|$ < 0.3, and $E\left(B - V\right)_{MW} < 0.15$.

\subsection{Estimating \prometal{}}
\label{subsec:z_method}

\citetalias{Kim2024} employed YEPS as their main results, as discussed in \citet{Kang2020}.
Accordingly, this work also used [Fe/H] and [$\alpha$/Fe] estimated from the YEPS model.
Thus, to determine the total metallicity ($Z$), we adopted the conversion among $Z$, [Fe/H], and [$\alpha$/Fe] from the YEPS model paper \citet{Chung2013} (see their Section 2.2);

\begin{equation}
\label{eq:z_feh_afe}
\text{log}(\frac{Z}{Z_{\odot}}) \text{ = }  \left[ Z \text{/} H \right] \text{ = }  \left[ Fe \text{/} H \right] + 0.723\left[ \alpha \text{/} Fe \right]
\end{equation}

From this equation, we determined the total metallicity of the birth environments and, thus, of the SN Ia progenitor star for our sample.
We plot the correlation between $Z$, [Fe/H] and [$\alpha$/Fe] of the progenitor stars in Fig.~\ref{fig:proz_feh_afe}.

\subsection{Estimating \nickel{} mass}
\label{subsec:nimass_method}

\nickel{} mass synthesized during the SN Ia explosion can be inferred from the SN Ia bolometric luminosity and its rise time using Arnett's Rule \citep{Arnett1979, Arnett1982}.
\citet{Scalzo2019} determined the \nickel{} mass from the SN Ia bolometric light-curve and then provided a fitting formula in terms of a SALT2 light-curve shape parameter $x_1$:

\begin{equation}
\label{eq:ni_x1}
$\nickel{}$ \text{ mass} \text{ = }  (0.659 \pm 0.023) + (0.136 \pm 0.019)x_{1}
\end{equation}

We employed this equation to obtain \nickel{} mass from SALT2 $x_{1}$ for our sample.

For estimating an uncertainty for each value, we used the python package \texttt{uncertainties}\footnote{\href{https://pythonhosted.org/uncertainties/}{https://pythonhosted.org/uncertainties/}}.

All data used in this work are available on the GitHub webpage\footnote{\href{https://github.com/Young-Lo/SNIaProgenitorZvsNiMass}{https://github.com/Young-Lo/SNIaProgenitorZvsNiMass}}.

\begin{table*}
\centering
\caption{Summary of the SN Ia explosion models employed in the present work. }
\label{tab:simul_summary}
\begin{tabular}{l c c c c c c c}
\hline\hline\\[-0.8em]
Ref.  & Model  &  Mechanism   & Model Name  &  $\rho_{c}$ 			& $M_{WD}$ 	& $M_{He}$ 		& \prometal{}   \\
        &	       &  		       & [used in the present work]		        &  [$10^8$ g cm$^{-3}$]  	& [\solarmass{}] 		& [\solarmass{}] 			& [$Z_{\odot}$]   \\
\hline \\ [-0.5em]
\multirow{3}{*}{\citetalias{Leung2018}}   & \multirow{3}{*}{Near-\chmass{}}  	&  \multirow{3}{*}{DDT} 	& \citetalias{Leung2018}\_100 & 10 & 1.33 & -- & \multirow{3}{*}{0, 0.1, 0.5, 1, 2, 3, 5}   \\ [0.4em]
							   &								&					 & \citetalias{Leung2018}\_300 & 30 & 1.38 & -- &    \\ [0.4em]
							   &								&					 & \citetalias{Leung2018}\_500 & 50 & 1.39 & -- &    \\ [0.4em]
\hline \\ [-0.5em]
\multirow{3}{*}{\citetalias{Leung2020}}   & \multirow{3}{*}{Sub-\chmass{}}  	&  \multirow{3}{*}{DD} & \citetalias{Leung2020}\_110-100   & 0.62 & 1.10 & 0.10 & \multirow{3}{*}{0, 0.1, 0.5, 1, 2, 3, 5} \\ [0.4em]
							   &								&			  & \citetalias{Leung2020}\_110-050 & 0.62 & 1.10 & 0.05 &    \\ [0.4em]
							   &								&			  & \citetalias{Leung2020}\_100-050 & 0.32 & 1.00 & 0.05 &   \\ [0.4em]
\hline \\ [-0.5em]
&& \\ [0.4em]
\hline \hline \\ [-0.8em]
Ref.  & Model  &  Mechanism   & Model Name  &  $\rho_{c}$ 			& $M_{core}$ 	& $M_{shell}$ 		& \prometal{}   \\
        &	       &  		       & [used in the present work] 		        &  [$10^8$ g cm$^{-3}$]  	& [\solarmass{}] 		& [\solarmass{}] 			& [$Z_{\odot}$]   \\
\hline \\ [-0.5em]
\multirow{4}{*}{\citetalias{Gronow2021}} & \multirow{4}{*}{Sub-\chmass{}} & \multirow{4}{*}{DD} & \citetalias{Gronow2021}\_M11\_$XX$ & -- & 1.1 & 0.05 & \multirow{4}{*}{0.001, 0.1, 1, 3} \\ [0.4em]
 & & & \citetalias{Gronow2021}\_M10\_$XX$ & -- & 1.0 & 0.02, 0.03, 0.05, 0.1 &		\\ [0.4em]
& & & \citetalias{Gronow2021}\_M09\_$XX$ & -- & 0.9 &  0.03, 0.05, 0.1 	&		\\ [0.4em]
& & & \citetalias{Gronow2021}\_M08\_$XX$ & -- & 0.8 &	0.03, 0.05, 0.1 		&	\\ [0.4em]
\hline \\ 
\end{tabular}
 \tablefoot{
"Mechanism" is the SN Ia explosion mechanism: "DDT" means deflagration-to-detonation transition, and "DD" means double-detonation models.
"$\rho_c$" is the central density of the WD in the unit of $10^{8}$ g cm$^{-3}$.
"$M_{WD}$" and "$M_{He}$" represent the total mass of the WD and the helium envelope mass, respectively, in the unit of \solarmass{}.
"$M_{core}$" and "$M_{shell}$" indicate the initial core and the initial shell mass of the WD, respectively, in the unit of \solarmass{}.
"\prometal{}" is the progenitor metallicity in the unit of $Z_{\odot}$.
"Model Name" is the name of the model used in the present work based on the references.
For \citetalias{Leung2018}, we only consider $\rho_c$ (three digits), such that "100" means $\rho_c$ = 1.0 x $10^{9}$ g cm$^{-3}$.
For \citetalias{Leung2020}, the model name consists of $M_{WD}$ (three digits) and $M_{He}$ (three digits).
For \citetalias{Gronow2021}, the model name is based on the $M_{core}$ (two digits) and $M_{shell}$ (two digits), such that "M10\_05" means  $M_{core}$ = 1.0 \solarmass{} and  $M_{shell}$ =  0.05 \solarmass{}.
We note that "--" in the table indicates that no information was provided in the reference.
 }
\end{table*}

\subsection{Hydrodynamic simulations of SN Ia explosion models}
\label{subsec:snia_models}

In this section, we list and explain recent hydrodynamic simulations of SN Ia explosion models used in the present work.
We select three models, which provide a wider range of \prometal{}: \citet{Leung2018}, \citet{Leung2020}, and \citet{Gronow2021}.
Tab.~\ref{tab:simul_summary} presents a summary of them with model parameters.
We refer the reader to the original paper for more detailed discussions of the adopted input physics and simulations.

\subsubsection{\citet[hereafter \citetalias{Leung2018}]{Leung2018}}
\label{subsubsec:ln18}

\citetalias{Leung2018} presented 2D hydrodynamic simulations of near-\chmass{} WD models using the deflagration-to-detonation transition (DDT) mechanism.
They explored a wide range of the parameter space for 41 models to study the effect of the initial $\rho_{c}$ (i.e., WD mass ($M_{WD}$)), \prometal{}, flame shape, DDT criteria, and turbulent flame formula.
The parameter space includes SNe Ia models with $\rho_{c}$ (from $0.5 \times 10^{9}$ g cm$^{-3}$ to $5 \times 10^{9}$ g cm$^{-3}$; $M_{WD}$ from 1.30 to 1.38 \solarmass{}), \prometal{} (from 0 to 5 \solarmetal{}).
SNe Ia yields from $^{11}$C to $^{91}$Tc are obtained by post-process nucleosynthesis calculations.

In the present work, we consider three models at seven different metallicities (\prometal{} = 0, 0.1, 0.5, 1, 2, 3, and 5 \solarmetal{}), for which tables are provided.
The models are the low-density ($\rho_{c}$ = $10 \times 10^{8}$ g cm$^{-3}$, $M_{WD}$ = 1.33 \solarmass{}; the model name "100" for the present work), the benchmark model ($\rho_{c}$ = $30 \times 10^{8}$ g cm$^{-3}$, $M_{WD}$ = 1.38 \solarmass{}; the model name "300"), and the high-density model ($\rho_{c}$ = $50 \times 10^{8}$ g cm$^{-3}$, $M_{WD}$ = 1.39 \solarmass{}; the model name "500").

\subsubsection{\citet[hereafter \citetalias{Leung2020}]{Leung2020}}
\label{subsubsec:ln20}

Using the same simulation as \citetalias{Leung2018}, \citetalias{Leung2020} extended their parameter survey of the SN Ia models to the double-denotation (DD) explosions of sub-\chmass{} WD.
They studied the effect of different \prometal{} (from 0 to 5 \solarmetal{}), $M_{WD}$ (from 0.90 to 1.20 \solarmass{}), He envelope masses ($M_{He}$; from 0.05 to 0.20 \solarmass{}), and the geometry of the He detonation (spherial, bubble, and ring).

In the present work, we use three models at seven different metallicities (\prometal{} = 0, 0.1, 0.5, 1, 2, 3, and 5 \solarmetal{}), for which tables are provided.
We take models of "110-050-2-B50", "110-100-2-50" (the benchmark model), and "100-050-2-S50", as named in \citetalias{Leung2020}.
The model name consists of $M_{WD}$ (three digits), $M_{He}$ (three digits), \prometal{} (one digit), and the initial position of the detonation bubble (two digits), with an additional one-character "B" or "S", which stand for different initial He detonations.
\citetalias{Leung2020} explained that the term "S50" stands for a spherical detonation triggered at 50 km above the He/CO interface, and "B50" stands for a belt (ring) around the "equator" of the WD.
For simplicity, we use the first six digits for the model name in the present work, as shown in Tab.~\ref{tab:simul_summary}.

\subsubsection{\citet[hereafter \citetalias{Gronow2021}]{Gronow2021}}
\label{subsubsec:g21} 

\citetalias{Gronow2021} presented 3D hydrodynamic simulations of sub-\chmass{} WD models using the DD explosions.
They computed and analyzed a set of 11 different models with varying WD core masses ($M_{core}$; from 0.8 to 1.1 \solarmass{}) and He shell masses ($M_{shell}$; from 0.02 to 0.1 \solarmass{}) at four different metallicities (\prometal{} = 0.001, 0.1, 1, and 3 \solarmetal{}) each.
SNe Ia yields consisting of 384 isotopes are obtained by post-processing with an extended nuclear network.

In the present work, we use all 11 models at four different metallicities, as \citetalias{Gronow2021} provided all the results.
The model name for the present work is based on $M_{core}$ (two digits) and $M_{shell}$ (two digits).
For example, "M10\_05" refers to all models with $M_{core}$ = 1.0 \solarmass{} and $M_{shell}$ = 0.05 \solarmass{}, thereby combining four models at different metallicity.

\section{Result}
\label{sec:result}

\begin{figure*}
	\centering	
	\sidecaption
	\includegraphics[width=12cm]{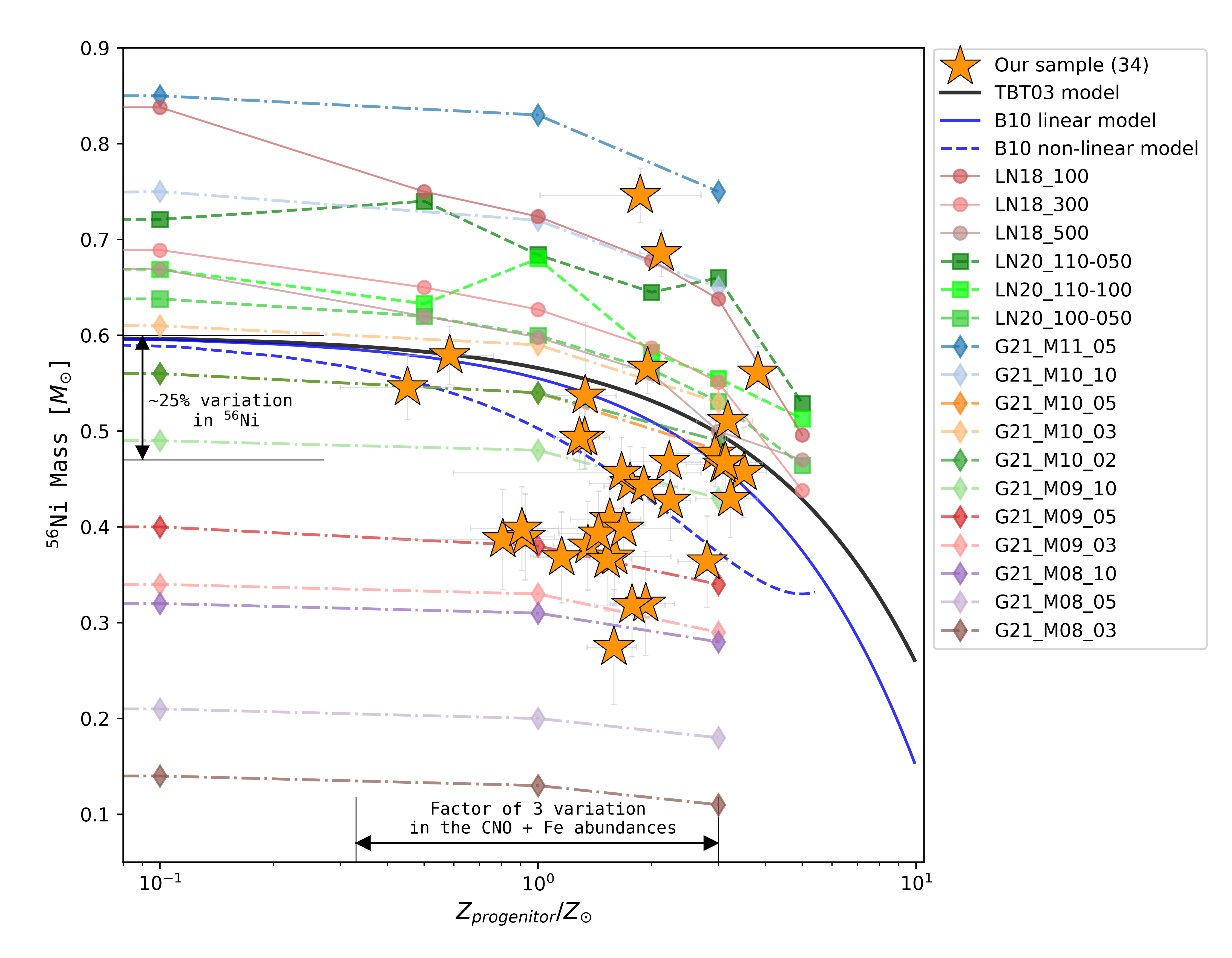}
	\caption{
	\nickel{} mass synthesized during the SN Ia explosion as a function of the progenitor metallicity (\prometal{}).
	Our sample of the 34 normal SNe Ia are indicated with star-shaped orange marks.
	 The analytical \citetalias{Timmes2003} (a black solid curved line) and \citetalias{Bravo2010} linear and non-linear (blue solid and blue dashed lines, respectively) models are overplotted with various SN Ia models investigated in the present work: 3 LN18 near-\chmass{} models (circles with solid lines), 3 LN20 sub-\chmass{} models (squares with dashed lines), and 11 G21 sub-\chmass{} models (diamonds with dash-dotted lines).
	 Our sample is well-distributed across the \prometal{} range where most of the \prometal{} effect occurs, as indicated by notes taken from figure 1 of \citetalias{Timmes2003}.
	 The derived \nickel{} masses of our sample are within the \nickel{} mass range expected from various hydrodynamic simulations.
	} 
	\label{fig:nimass_z_all}
\end{figure*}

We first investigate the \nickel{} mass as a function of \prometal{}, focusing on the anlytical \citetalias{Timmes2003} and \citetalias{Bravo2010} predictions.
Then, we try to constrain the SN Ia progenitors by comparing our sample with hydrodynamic simulations on the \nickel{} mass and \prometal{} diagram.
We further study an impact of progenitor's [Fe/H] and [$\alpha$/Fe] on \nickel{} mass synthesized during the SN Ia explosion.
Lastly, a result of \prometal{} versus the corrected luminosity of SNe Ia will be presented.

\subsection{Constraining SN Ia progenitors on the \prometal{}--\nickel{} mass diagram}
\label{subsec:ni_vs_z}

Fig.~\ref{fig:nimass_z_all} shows our main result: \nickel{} mass synthesized during the SN Ia explosion as a function of \prometal{}.
We have plotted the \citetalias{Timmes2003} model with a black solid line and notes taken from their figure 1.
\citetalias{Bravo2010} linear (blue solid line) and non-linear (blue dashed line) models are also presented.
For these models, we set that a fiducial SN Ia produces $\approx 0.6M_{\odot}$ of \nickel{} as in \citetalias{Timmes2003}. 
All the results from hydrodynamic simulations of SN Ia models are shown with different colours and marks.

Compared to previous studies, which only have a sample in the sub-solar range, our sample covers the entire \prometal{} range explored by \citetalias{Timmes2003}: $\frac{1}{3}$ $Z_{\odot}$ < \prometal{} < 3 $Z_{\odot}$.
In this range, most of the \prometal{} effect occurs.
Although there is a higher scatter in the derived \nickel{} masses compared to the analytical predictions, our sample is within the \nickel{} mass range expected from various hydrodynamic simulations.

All hydrodynamic simulations and models of \citetalias{Bravo2010} show the same trend with \citetalias{Timmes2003}, but with a different size of \nickel{} mass variation.
\citetalias{Bravo2010} shows larger variation in both linear and non-linear models, because they found a steeper slope than \citetalias{Timmes2003}.
\citetalias{Leung2018} near-\chmass{} models show a similar variation: $\sim$24\%.
However, sub-\chmass{} models show less variation, but similar sizes between them: $\sim$15\% and $\sim$17\% for \citetalias{Leung2020} and \citetalias{Gronow2021}, respectively.

\begin{figure}
	\centering	
	\includegraphics[width=\columnwidth]{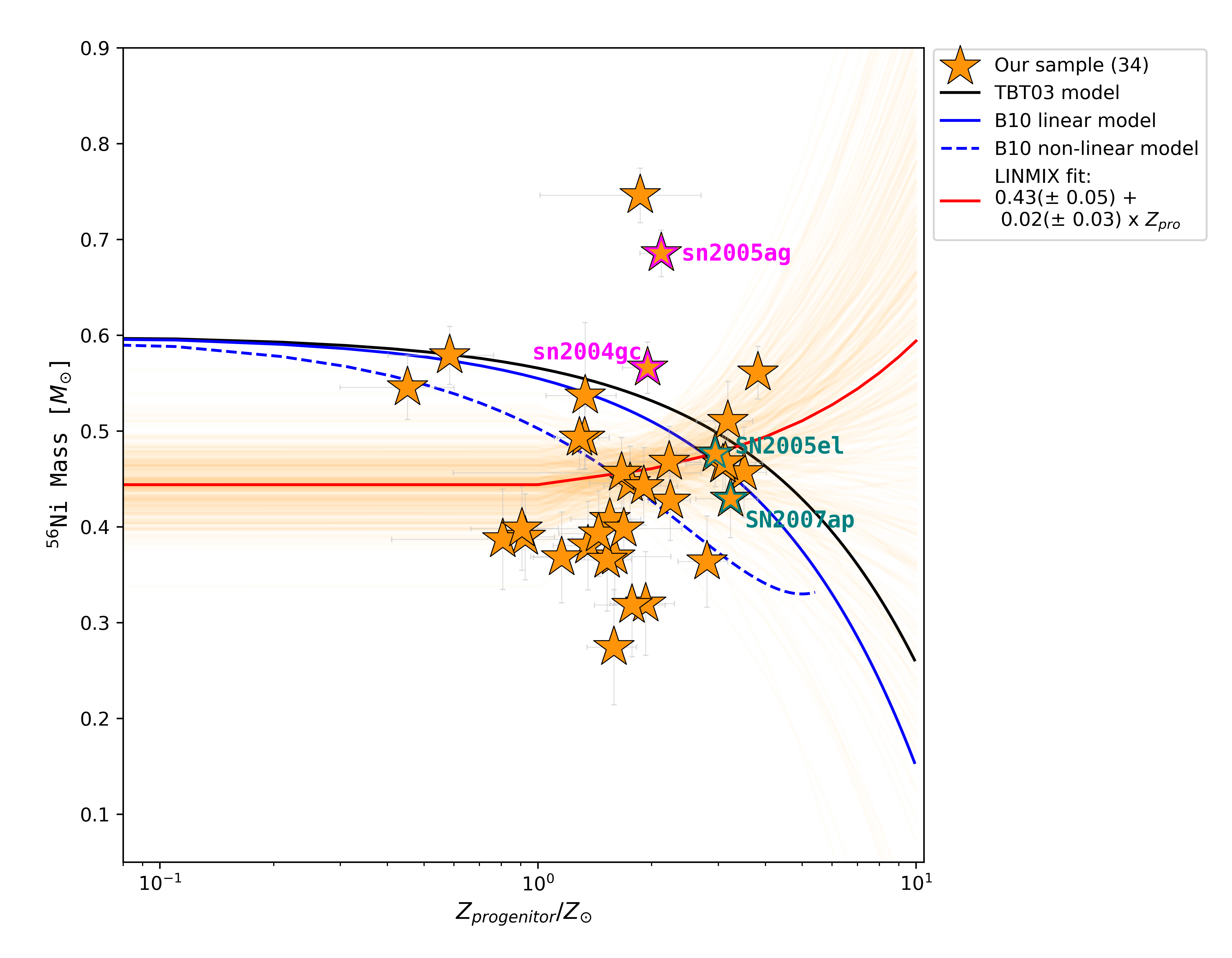}
	\includegraphics[width=\columnwidth]{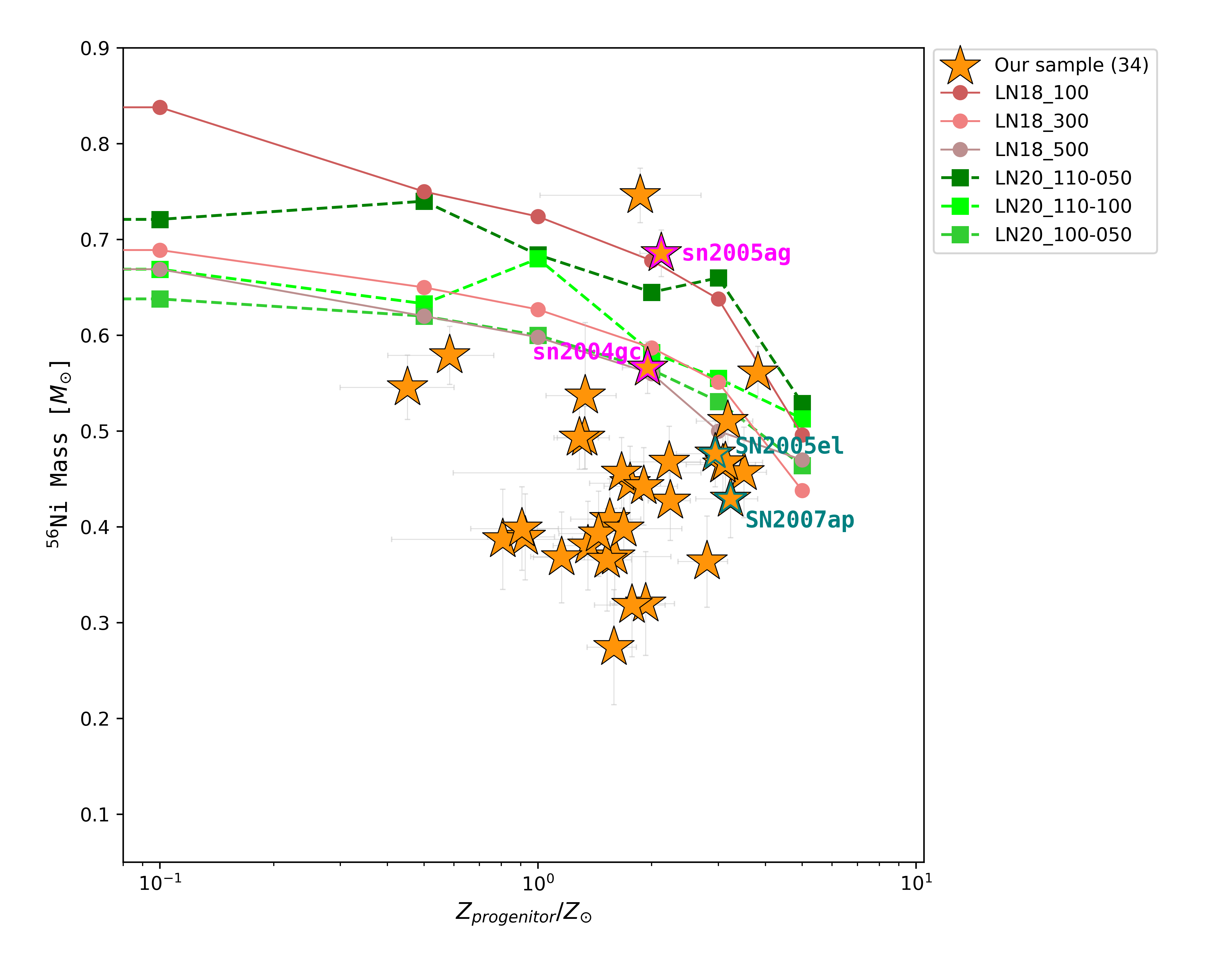}
	\includegraphics[width=\columnwidth]{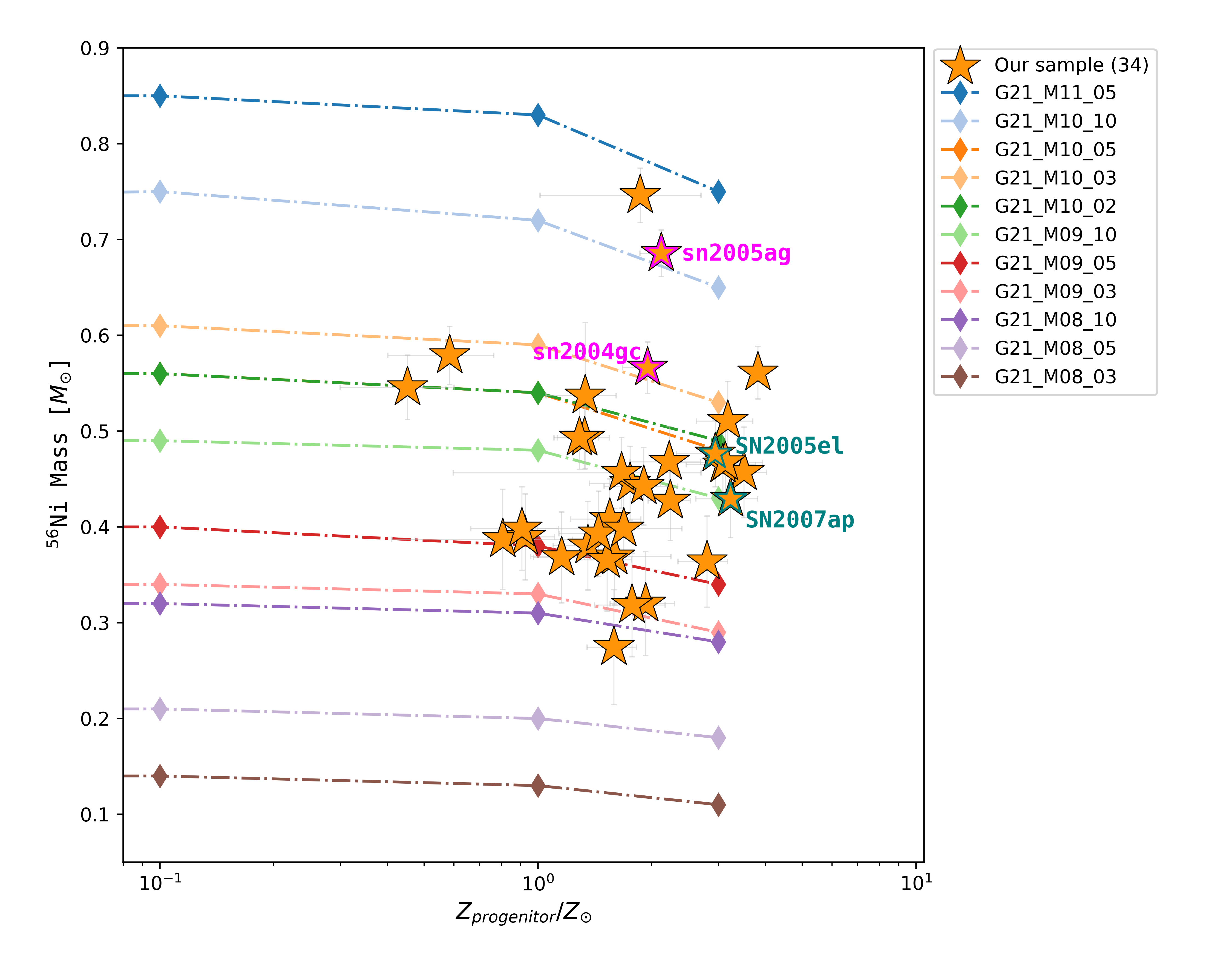}
	\caption{
	Same as Fig.~\ref{fig:nimass_z_all}, but split the plot by the SN Ia models.
	The top panel is for \citetalias{Timmes2003} and \citetalias{Bravo2010}.
	A red solid line shows the average of 10,000 linear regression results (light red lines) from the LINMIX package, which returns a slope of $0.02 \pm 0.03$ (0.7$\sigma$).
	The middle panel is for \citetalias{Leung2018} and \citetalias{Leung2020} models, and the bottom panel is for the \citetalias{Gronow2021} model.
	In each panel, we mark four SNe Ia, which show a good fit with \citetalias{Leung2018} and \citetalias{Leung2020} models in magenta colour and with the \citetalias{Gronow2021} model in teal colour.
	} 
	\label{fig:nimass_z_split}
\end{figure}

In the following sections and Fig.~\ref{fig:nimass_z_split}, we investigate in detail split by models.

\subsubsection{\citetalias{Timmes2003} and \citetalias{Bravo2010} models}
\label{subsubsec:tbt03_b2010}

Here, we investigate the relation between \prometal{} and the \nickel{} mass rather than constraining the progenitors, as discussed in \citetalias{Timmes2003} and \citetalias{Bravo2010} (the top panel of Fig.~\ref{fig:nimass_z_split}).
Although some of our sample of SNe Ia follow the trend explored by \citetalias{Bravo2010} linear and non-linear models, the scatter in the determined \nickel{} mass is higher than those predicted by models, as discussed in previous studies (e.g., \citetalias{Howell2009} and \citealt{Neill2009}).
Since \citetalias{Timmes2003} and \citetalias{Bravo2010} provided equations between \prometal{} and \nickel{} mass, we derived an equation from our sample to compare them.
To do this, we perform 10,000 linear regressions on our data using the LINMIX package in $Python$\footnote{\href{https://github.com/jmeyers314/linmix/}{https://github.com/jmeyers314/linmix/}}, which employs an MCMC posterior sampling and a hierarchical Bayesian approach, considering errors in both variables \citep{Kelly2007}.
The LINMIX result shows that the slope of our data is $0.02\pm0.03$ with low statistical significance at 0.7$\sigma$.
This appears to be the opposite trend to \citetalias{Timmes2003} and \citetalias{Bravo2010} predictions, but the slope is statistically consistent with zero.
We caution that the \prometal{} range ($\frac{1}{2}$ \solarmetal{} < \prometal{} < 4 \solarmetal{}) is too narrow to determine a statistically meaningful result.

\subsubsection{\citetalias{Leung2018} and \citetalias{Leung2020} models}
\label{subsubsec:ln2018_2020}

At first glance at the bottom left panel of Fig.~\ref{fig:nimass_z_split}, we see a lack of models in the \nickel{} mass range lower than 0.6 \solarmass{}.
This might be because \citetalias{Leung2018} and \citetalias{Leung2020} selected their benchmark models as ones that can produce a \nickel{} mass of around 0.6 \solarmass{}, considering observations from normal SNe Ia.
When we check their model setup for benchmark models (table 2 of \citetalias{Leung2018} and table 1 of \citetalias{Leung2020}), we found that SNe Ia with \nickel{} mass between 0.2 and 0.4 \solarmass{} can be found when considering only deflagration for the near-\chmass{} model\footnote{It is worth to note that models with deflagration only are aimed at reproducing the peculiar Type Iax SNe class \citep[e.g.,][]{Kromer2015}.} or 0.9 \solarmass{} with "Y" type explosion for the sub-\chmass{} model.

In the figure, we mark two SNe Ia in magenta that appear to be well constrained by the model predictions.
SN 2005ag is well matched with "\citetalias{Leung2018}\_100" at \prometal{} = 2 \solarmetal{} model.
It suggests that its progenitor can be a WD with a total mass of 1.33 \solarmass{} and \prometal{} = 2 \solarmetal{} exploded via the near-\chmass{} DDT mechanism.
SN 2004gc could be explained by two models: "\citetalias{Leung2018}\_500" at 2 \solarmetal{} and "\citetalias{Leung2020}\_100-050" at 2 \solarmetal{}.
Additional information, such as $^{57}$Ni mass, which can be obtained from SN Ia observations (see e.g., \citealt{Dimitriadis2017}), is required to distinguish between models.

\subsubsection{\citetalias{Gronow2021} models}
\label{subsubsec:g2021}

In the bottom right panel of Fig.~\ref{fig:nimass_z_split}, \citetalias{Gronow2021} models cover a wide range of the \nickel{} mass and show a consistent trend with \prometal{} between them.
Our sample is well distributed within the range covered by the models. 
In particular, many of our SNe Ia are between "\citetalias{Gronow2021}\_09\_10" (light green) and "\citetalias{Gronow2021}\_09\_05" (red) at \prometal{} > 1 \solarmetal{} models.

We mark in teal two SNe Ia that that appear to be well constrained by the \citetalias{Gronow2021} model.
SN 2005el shows a good fit with "\citetalias{Gronow2021}\_M10\_05" at \prometal{} = 3 \solarmetal{} model, while SN 2007ap can be explained by "\citetalias{Gronow2021}\_M09\_10" at \prometal{} = 3 \solarmetal{} model.
Their progenitors are WDs with the same \prometal{} = 3 \solarmetal{}, but different total masses of 1.05 \solarmass{} ( 1.0 \solarmass{} of $M_{core}$ +  0.05 \solarmass{} of $M_{shell}$) and 1.0 \solarmass{} (0.9 \solarmass{} of $M_{core}$ + 0.10 \solarmass{} of $M_{shell}$), respectively, exploded via the sub-\chmass{} DD mechanism.

We note that the "\citetalias{Gronow2021}\_M10\_05" model (orange) and "\citetalias{Gronow2021}\_M10\_02" model (green) almost overlap. 
They show just 0.01 \solarmass{} difference in the \nickel{} mass at \prometal{} = 3 $Z_{\odot}$.
This might be due to the different configurations of the models, but a detailed analysis of this is beyond the scope of the present work.

\subsubsection{SN Ia progenitors from different models}
\label{subsubsec:progenitors}

\begin{table*}
\centering
\caption{Summary of the SN Ia progenitors constrained in the present work. }
\label{tab:pro_summary}
\begin{tabular}{l c c c c c c c}
\hline\hline\\[-0.8em]
SN &  Model Name  			& Model &  Mechanism  & $M_{WD}$ 	& $M_{core}$ 		& $M_{shell}$/$M_{He}$ &  \prometal{}   \\
       & [used in the present work]	&	       &  		      & [\solarmass{}] 	& [\solarmass{}] 	& [\solarmass{}] 		&	[$Z_{\odot}$]   \\
\hline \\ [-0.5em]
2007ap 				& \citetalias{Gronow2021}\_M09\_10 & Sub-\chmass{} 	& DD 	& 1.00 	& 0.9 	& 0.1 	& 3 		\\[0.4em] 
\hline \\ [-0.5em]
\multirow{2}{*}{2005ag}	& \citetalias{Leung2018}\_100 		& Near-\chmass{}	& DDT	& 1.33	& --		& --		& 2  		\\ [0.4em]
				   	& \citetalias{Gronow2021}\_M10\_10	& Sub-\chmass{}	& DD	& 1.10	& 1.0		& 0.1		& 1-3	\\ [0.4em]
\hline \\ [-0.5em]
\multirow{2}{*}{2005el}	& \citetalias{Gronow2021}\_M10\_05	& Sub-\chmass{}	& DD	& 1.05	& 1.0		& 0.05	& 3		\\ [0.4em]
				 	& \citetalias{Leung2018}\_500 		& Near-\chmass{}	& DDT	& 1.39	& --		& --		& 3  		\\ [0.4em]
\hline \\ [-0.5em]
\multirow{3}{*}{2004gc} 	& \citetalias{Leung2018}\_500 		& Near-\chmass{}	& DDT	& 1.39	& --		& --		& 1  		\\ [0.4em]
				    	& \citetalias{Leung2020}\_100-050 	& Sub-\chmass{}	& DD	& 1.00	& 0.95	& 0.05	& 1  		\\ [0.4em]
				    	& \citetalias{Gronow2021}\_M10\_03	& Sub-\chmass{}	& DD	& 1.03	& 1.0		& 0.03	& 1-3	\\ [0.4em]
\hline 
\end{tabular}
\end{table*}

In the previous sections, for illustrative purposes, we present candidates for progenitors of four SNe Ia in our sample by comparing them to the hydrodynamic simulations of SN Ia models on the \nickel{} mass--\prometal{} diagram.
Other SNe Ia in the sample are also in close agreement with the model predictions by \citetalias{Leung2018}, \citetalias{Leung2020}, and \citetalias{Gronow2021}. 
We mainly selected one model that was considered to show a good fit between the data and the model.
However, it seems that the above four SNe Ia can also be explained by other models, which have different configurations.
Thus, we try to consider additional models in this section.
Tab.~\ref{tab:pro_summary} shows the summary of them.

\begin{description}
\item[-\textit{SN 2007ap}:]
In the present work, only the "\citetalias{Gronow2021}\_M09\_10" at 3 \solarmetal{} model can explain this SN.
However, a pure deflagration explosion of a near-\chmass{} WD of \citetalias{Leung2018} or a DD explosion of 0.9 \solarmass{} WD of \citetalias{Leung2020} may also be a progenitor of this SN Ia.
\end{description}
\begin{description}
\item[-\textit{SN 2005ag}:]
The "\citetalias{Leung2018}\_100" at 2 \solarmetal{} model shows a good fit with this SN Ia.
We see that SN 2005ag is close to the line connecting the "\citetalias{Gronow2021}\_M10\_10" at 1 \solarmetal{} model and that at 3 \solarmetal{}.
This means that a progenitor of SN 2005ag also can be determined by  "\citetalias{Gronow2021}\_M10\_10" at 2 \solarmetal{}.
\end{description}
\begin{description}
\item[-\textit{SN 2005el}:]
The "\citetalias{Gronow2021}\_M10\_05" at 3 \solarmetal{} model can best describe this SN Ia.
The "\citetalias{Leung2018}\_500" at 3 \solarmetal{} model could have a chance to explain this SN, too.
\end{description}
\begin{description}
\item[-\textit{SN 2004gc}:]
For this SN Ia, we selected two models: "\citetalias{Leung2018}\_500" at 2 \solarmetal{} and "\citetalias{Leung2020}\_100-050" at 2 \solarmetal{}.
SN 2004gc is on the line connecting the "\citetalias{Gronow2021}\_M10\_03" at 1 \solarmetal{} model and that at 3 \solarmetal{}.
\end{description}

This section shows that additional observables from SNe Ia or host galaxies to distinguish between models are required, as discussed above section.
Along with this observational effort, hydrodynamic simulations of SN Ia models with denser grids in the parameter space are needed to fill the gaps between the current models.

\subsection{\nickel{} mass as a function of [Fe/H] and [$\alpha$/Fe]}
\label{subsec:ni_vs_feh_afe}

\begin{figure*}
	\centering	
	\includegraphics[width=0.8\textwidth]{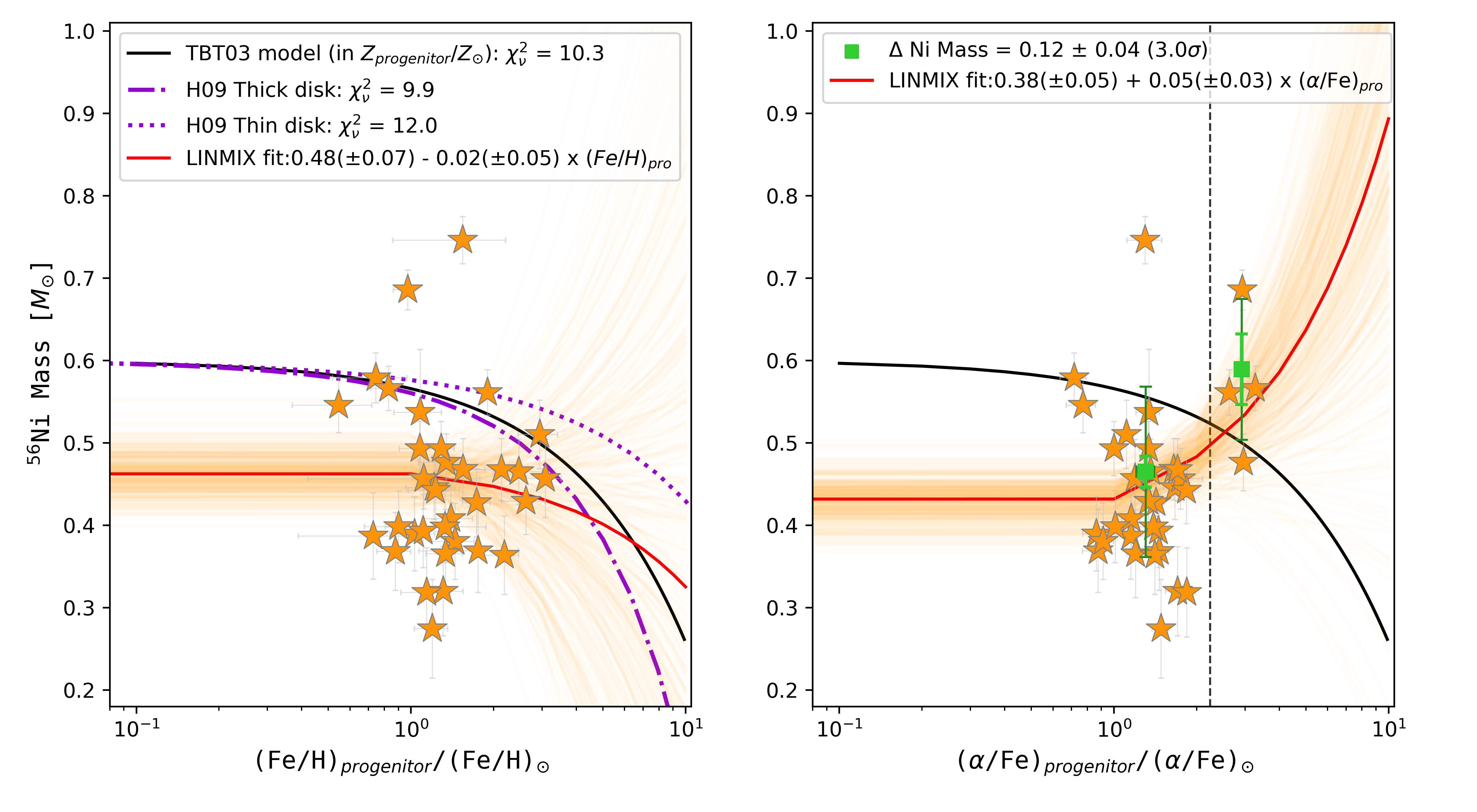}
	\caption{
	\nickel{} mass synthesized during the SN Ia explosion as a function of the progenitor iron abundance ($(Fe/H)_{progenitor}$) and $\alpha$-element enrichment ($(\alpha/Fe)_{progenitor}$).
	The \citetalias{Timmes2003} model is plotted as a solid black line, even in the expression of \prometal{} / $Z_{\odot}$.
	Red solid lines show the average of 10,000 linear regression results (light red lines) from the LINMIX package.
	In the left panel, purple dot-dashed and dotted lines present the \citetalias{Timmes2003} model, altered for thick and thin disk models by \citetalias{Howell2009}, respectively.
	A vertical dashed line in the right panel indicates \proafe{} = 0.35, showing a clear gap as discussed in \citetalias{Kim2024}.
	Green squares represent the weighted-means of  \nickel{} mass in each $(\alpha/Fe)_{progenitor}$ bin.
	Our sample seems to favour the \citetalias{Howell2009} thick disk model in terms of the relation between $(Fe/H)_{progenitor}$ and \nickel{} mass, given the $\chi^{2}_{\nu}$ value.
	The opposite trend to the \citetalias{Timmes2003} model for $(\alpha/Fe)_{progenitor}$ is observed.
	} 
	\label{fig:nimass_afe_feh}
\end{figure*}

\begin{table*}
\centering
\caption{The weighted-means of  \nickel{} mass and Hubble residuals in the different environments.}
\label{tab:stat}
\begin{tabular}{l  c c | c c c}
\hline\hline\\[-1.em]
(the right panel of Fig.~\ref{fig:nimass_afe_feh}) & $N_{SN}$  & \nickel{} mass              &    (Fig.~\ref{fig:nimass_hr})    & $N_{SN}$  & Hubble residual  \\
 &                   & [$M_{\odot}$]			      &				    &			& [mag]  \\
\hline \\ [-0.9em]
\proafe-rich   & 4  	& $0.59 \pm 0.04$  & \prometal-rich    & 8                & $0.15 \pm 0.09$     \\ [0.4em]
\proafe-poor & 30  	& $0.47 \pm 0.02$   &\prometal-poor   & 26              & $0.01 \pm 0.03$    \\ [0.4em]
\hline
Difference    &    	& $0.12 \pm 0.04$ (3.0$\sigma$) 	   &&			& $0.14 \pm 0.09$ (1.6$\sigma$)   \\
\hline
\end{tabular}
\end{table*}

As described in Sec.~\ref{sec:method}, \prometal{} is determined from the combination of [Fe/H] and [$\alpha$/Fe] via Eq.~\ref{eq:z_feh_afe}.
Thus, in order to see whether [Fe/H] and [$\alpha$/Fe] also have a trend with the \nickel{} mass, we plot \nickel{} mass as a function of [Fe/H] and [$\alpha$/Fe] in Fig.~\ref{fig:nimass_afe_feh}.
The \citetalias{Timmes2003} model is shown in the figure, although the model is described in \prometal{}.
Furthermore, we have overplotted the updated \citetalias{Timmes2003} model by \citetalias{Howell2009}.
As explained in Sec.~\ref{sec:intro}, \citetalias{Howell2009} accounted for the fact that $O/Fe$ can vary as a function of $Fe/H$ and different populations of stars in the thin disk, the thick disk, and the halo.
In the figure, we use Eq.~\ref{eq:h09_feh}, which describes the correlation between the \nickel{} mass and [Fe/H].
We note that we put x-axis labels as $(Fe/H)_{progenitor}$ / $(Fe/H)_{\odot}$ and $(\alpha/Fe)_{progenitor}$ / $(\alpha/Fe)_{\odot}$, instead of [Fe/H] and [$\alpha$/Fe] to match that of figures above.

In the relation between $(Fe/H)_{progenitor}$ and \nickel{} mass (the left panel of Fig.~\ref{fig:nimass_afe_feh}), we obtain a slope of $-0.02\pm0.05$ (0.4$\sigma$) for our sample using the LINMIX package.
This is the same trend as \citetalias{Timmes2003} and its updated models, but the magnitude of the slope is smaller and even consistent with zero when considering the error of the slope.
Then, we quantify the goodness-of-fit based on the reduced $\chi^{2}$ ($\chi^{2}_{\nu}$) statistic.
Our sample appears to prefer the thick disk model ($\chi^{2}_{\nu}$ = 9.9) to the thin disk model ($\chi^{2}_{\nu}$ = 12.0), and the \citetalias{Timmes2003} model ($\chi^{2}_{\nu}$ = 10.3).

Noticeably, for $(\alpha/Fe)_{progenitor}$, our sample shows the opposite trend to that of the \citetalias{Timmes2003} model, but note that this model is expressed in \prometal{}.
We split the sample into two groups based on \proafe{} = 0.35, showing a clear gap, and calculated weighted-means of \nickel{} mass in each \proafe{} bin.
The \nickel{} mass difference in the different \proafe{} groups is $0.12\pm0.04$ (3.0$\sigma$) $M_{\odot}$ (see Tab.~\ref{tab:stat}).
Linear regression using the LINMIX gives the slope of $0.05\pm0.03$ (1.7$\sigma$). 
We caution again that the ranges of $(Fe/H)_{progenitor}$ / $(Fe/H)_{\odot}$ and $(\alpha/Fe)_{progenitor}$ / $(\alpha/Fe)_{\odot}$ are too narrow to determine the statistically meaningful result for the LINMIX linear regression.
However, we can say that the difference in weighted-means of \nickel{} mass in the different \proafe{} groups is statistically significant at the 3$\sigma$ confidence level.

\subsection{\prometal{} versus the Hubble residual}
\label{subsec:z_vs_hr}

\begin{figure}
	\centering	
	\includegraphics[width=\columnwidth]{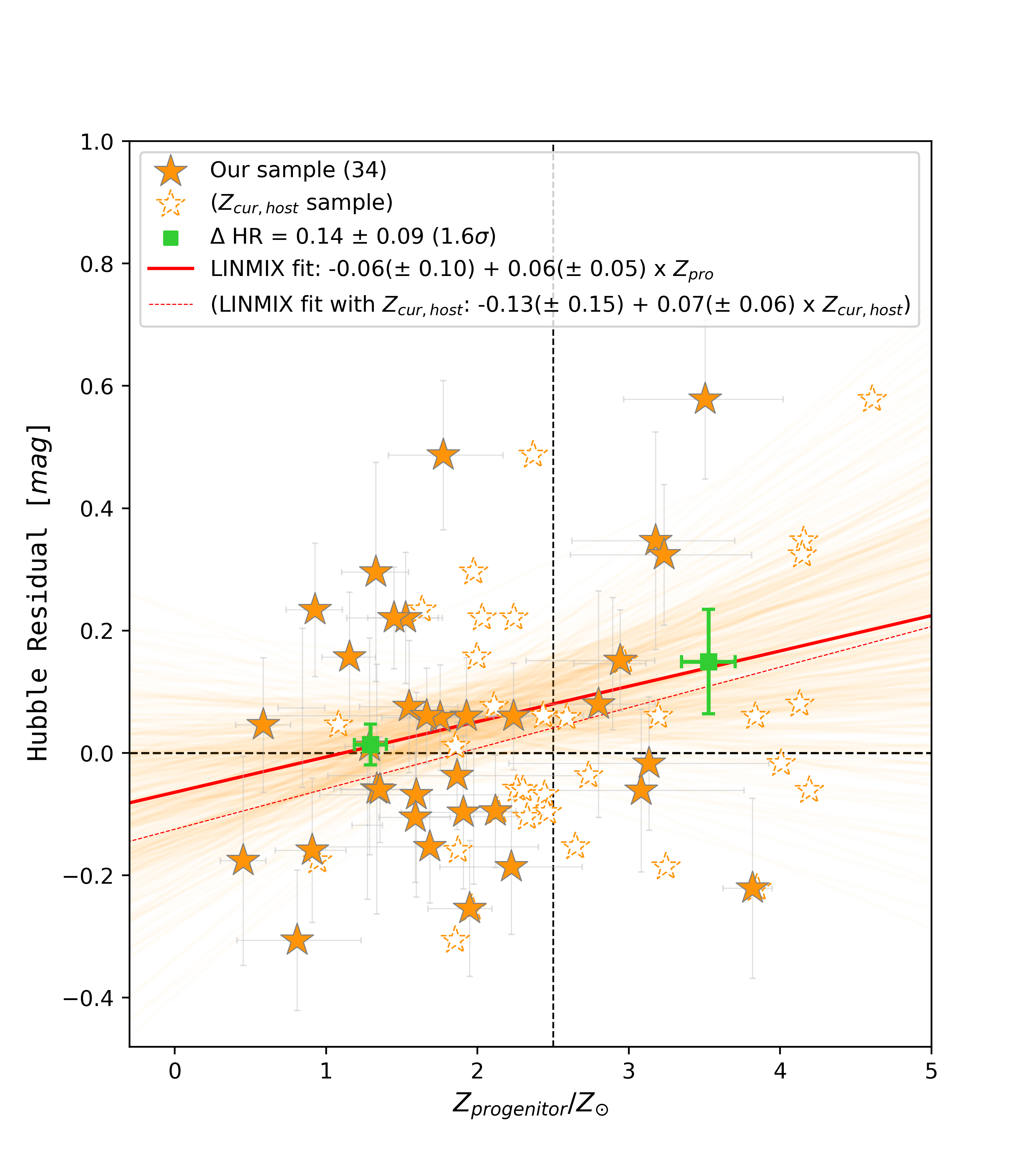}
	\caption{
	Impact of \prometal{} on the SN Ia corrected luminosity (i.e. the Hubble residual).
	Samples with $Z$ of current host galaxies ($Z_{cur,host}$) are also presented with empty star marks.
	The average of 10,000 linear regression results (light red lines) from the LINMIX package is shown with a red solid line.
	The LINMIX fit with $Z_{cur,host}$ is indicated with a red dashed line.
	We split our sample at \prometal{} = 2.5 (the vertical dashed line), where a gap is shown.
	Green squares are weighted-means of HR in the different \prometal{} group.	
	The trend, although at a low significance level, is opposite to the previous studies using current host galaxy properties.
	} 
	\label{fig:nimass_hr}
\end{figure}

There are some previous studies, which investigated the dependence of a corrected SNe Ia luminosity, after the standard light-curve corrections, with the currently-observed gas-phase or stellar metallicity of host galaxies (e.g., \citealt{Gallagher2005}; \citetalias{Howell2009}; \citealt{Childress2013}; \citealt{Pan2014}; \citealt{MorenoRaya2018}; \citealt{Kang2020}; \citealt{Galbany2022}; \citealt{MillanIrigoyen2022}).
In contrast to the well-established host mass-step \citep{Kelly2010, Lampeitl2010, Sullivan2010}, it is not yet well-established whether a dependence of SNe Ia luminosity on the host galaxy metallicity exists.
\citetalias{Kim2024} discussed that the environmental dependence studies using the currently-observed status of host galaxies could result in a weak correlation between SN Ia properties and host galaxy properties, because 1) the currently-observed status of host galaxies typically used in previous studies is different from the true SN Ia progenitor star birth environments and 2) the range of the host galaxy properties is narrower than that of the birth environments of the SN Ia progenitor stars.
Because the analytical prediction of \citetalias{Timmes2003} is performed with \prometal{} and presented 0.2 mag variation in the intrinsic luminosity of SNe Ia, it would be interesting to see the dependence of a corrected SNe Ia luminosity with \prometal{}.

In Fig.~\ref{fig:nimass_hr}, we investigate the impact of \prometal{} on the corrected luminosity of SNe Ia, called the Hubble residual (HR $\equiv$ $\mu_{SN} - \mu_{z}$).
We note that we do not include a mass-correction term in $\mu_{SN}$ and an intrinsic scatter term when calculating HRs.
This is because we want to examine what physical property could vary the SN Ia luminosity.
For the HR calculation, $\Omega_{M}$ = 0.3 and $H_0$ = 70 $km \: s^{-1} \: Mpc{-1}$ are used assuming flat $\Lambda$CDM cosmology.
We refer the reader to \citet{Kim2018, Kim2019} for details on the calculation of $\mu_{SN}$ and HR.

We performed 10,000 linear regressions for our data with the LINMIX package.
In addition, we split our sample into two \prometal{} groups at \prometal{} = 2.5, where a gap is shown, and then estimate the weighted-means of HRs in each group.
The slope of our data is $0.06\pm0.05$ ($1.2\sigma$) and the difference in the weighted-means of HRs is $0.14\pm0.09$ ($1.6\sigma$) mag (see Tab.~\ref{tab:stat}).
Both results show that SNe Ia from higher-metallicity progenitor stars are fainter than those from lower-metallicity progenitor stars after the light-curve corrections. 
This trend is inconsistent with the previous studies using current host galaxy properties (e.g., \citetalias{Howell2009}; \citealt{Pan2014}; \citealt{MillanIrigoyen2022}).
However, when we use $Z$ of current host galaxies (a red dashed line in the figure), we still see the opposite trend to previous studies, but the same trend with our \prometal{} result and also the same size in the \nickel{} mass difference but split the sample at 3.5 \solarmetal{}.

\section{Discussion and conclusions}
\label{sec:discussion}

In this work, we present the effect of the progenitor star property on the SN Ia intrinsic (i.e., \nickel{} mass) and corrected luminosity (i.e., HR).
For this, we try to reproduce the analytical model of \citetalias{Timmes2003} and its updated models by \citetalias{Howell2009} and \citetalias{Bravo2010}, which explored the correlation between the progenitor metallicity and the \nickel{} mass synthesized in SNe Ia during the explosion.
One step further, we attempt to constrain the SN Ia progenitors by comparing our data with recent SN Ia explosion simulations on this correlation.
Different from the previous studies that used the gas-phase metallicity inferred from the stellar mass of the host galaxies, we use 34 total metallicities of the birth environments of the normal SN Ia progenitor stars determined by \citetalias{Kim2024}, assuming this metallicity can be considered \prometal{}.

Our sample is well-distributed over the \prometal{} range indicated by \citetalias{Timmes2003} ($\frac{1}{3}$ $Z_{\odot}$ < \prometal{} < 3 $Z_{\odot}$), where most of the \prometal{} effect occurs, compared to previous studies, which only have a sample in the sub-solar range.
Our sample shows a higher scatter in the derived \nickel{} masses compared to those predicted by the analytical models.
Linear regression using the LINMIX package gives a slope of $0.02\pm0.03$ for our sample, which is the opposite trend to the \citetalias{Timmes2003} and \citetalias{Bravo2010} predictions.
However, the statistical significance of our slope is low at 0.7$\sigma$, and thus we can say that the slope is consistent with zero.

Then, we present three hydrodynamic simulations of SN Ia explosion models: the near-\chmass{} DDT model of \citetalias{Leung2018}, the sub-\chmass{} DD model of \citetalias{Leung2020}, and the sub-\chmass{} DD model of \citetalias{Gronow2021}.
All simulations show the same trend with \citetalias{Timmes2003}, but the magnitudes of \nickel{} mass variations are different.
Comparing our sample to those SN Ia explosion simulations on the \prometal{}--\nickel{} mass diagram, we try to constrain four progenitors, which show a good fit with the result from the simulations.
In the present work, SN 2007ap can be explained by only one model ("\citetalias{Gronow2021}\_M09\_10" at 3 \solarmetal{} model), and the other three SNe Ia (SN 2005ag, SN 2005el, and SN 2004gc) show good fits with two or three different models.
This means that additional observables of SNe Ia (e.g., $^{57}$Ni mass) and hydrodynamic simulations with denser grids in the parameter space are required to distinguish and to cover the gaps between the models.

We also investigate relations with progenitor's various other chemical abundances: $(Fe/H)_{progenitor}$ and $(\alpha/Fe)_{progenitor}$.
Regarding $(Fe/H)_{progenitor}$, our sample appears to follow the same trend as \citetalias{Timmes2003} and its updated models by  \citetalias{Howell2009}.
However, the magnitude of the slope is smaller: $-0.02\pm0.05$ (0.4$\sigma$) of our sample versus -0.057 of the analytical \citetalias{Timmes2003} model.
Our sample seems to favour the updated \citetalias{Timmes2003} model altered for thick-disk $O/Fe$ by \citetalias{Howell2009} ($\chi^{2}_{\nu}$ = 9.9), contrary to the previous discussion of \citetalias{Howell2009} and \citet{Neill2009}, which discussed in terms of the thin-disk model ($\chi^{2}_{\nu}$ = 12.0).
Our sample's preference for the thick disk is somewhat expected.
This is because the early-type galaxies used in the present work typically show the enhanced $[\alpha/Fe]$ \citep[e.g.,][]{Thomas2005}, as the thick disk exhibits a high $[\alpha/Fe]$ ratio relative to the thin disk \citep[e.g.,][and references therin]{Matteucci2021}.

Interestingly, with $(\alpha/Fe)_{progenitor}$, our sample shows the opposite trend to the \citetalias{Timmes2003} model.
The LINMIX package returns a slope of $0.05\pm0.03$ (1.7$\sigma$).
Because we see the gap at $(\alpha/Fe)_{progenitor}$ = 0.35, we split our sample into two groups and calculate the weighted-means of \nickel{} mass in each $(\alpha/Fe)_{progenitor}$ group. 
We found that the \nickel{} mass difference in the different \proafe{} groups is $0.12\pm0.04$ (3.0$\sigma$) $M_{\odot}$.

Lastly, we present the impact of \prometal{} on the corrected SN Ia luminosity.
When we split our sample into two \prometal{} groups, the difference in the corrected luminosity in each group is $0.14 \pm 0.09$ mag (1.6$\sigma$).
Linear regression shows that our sample has a slope of $0.05\pm0.06$ (0.8$\sigma$).
Our result is that SNe Ia from higher-metallicity progenitor stars are fainter than those from lower-metallicity progenitor stars, after the light-curve corrections (in terms of HRs).

If the analytical model is confirmed with more data, we expect that \prometal{} alone, which can vary the luminosity about 0.2 mag, cannot fully explain the observed scatter in the peak luminosity of SNe Ia (0.5 mag in $B$ and $V$-bands).
In addition, as suggested by the luminosity variation of SNe Ia in the different \prometal{} group ($0.14\pm0.09$ mag), the light-curve standardization process cannot fully account for the progenitor metallicity effect on the luminosity variation.
Through this standardization process, the luminosity variation is expected to decrease by $\sim$30\%, from 0.2 mag to 0.14 mag.
Other properties, such as the progenitor age, mass, and other chemical elements, will also contribute to the scatter.
Furthermore, different explosion scenarios, e.g. single-degenerate or double-degenerate, deflagration or detonation, near- or sub-\chmass{}, could produce a dispersion in the peak magnitudes of SNe Ia \citep[see][for reviews on SN Ia explosions]{Liu2023, Ruiter2024}.

Noticeably, $\alpha/Fe$ appears to play an interesting role in the SN Ia explosion.
In this work, the correlation between $(\alpha/Fe)_{progenitor}$ and the \nickel{} mass (the right panel of Fig.~\ref{fig:nimass_afe_feh}) shows the opposite trend to the \citetalias{Timmes2003} model, which is, however, expressed in \prometal{}.
In \citetalias{Kim2024}, $[\alpha/Fe]$ of the SN Ia progenitor star birth environment can clearly distinguish the SN Ia sample into two groups.
In the galaxy study, $[\alpha/Fe]$ is linked with the timescale of the star formation history ($\Delta t$), such that $[\alpha/Fe]$ $\sim$ $\frac{1}{5} - \frac{1}{6}log(\Delta t)$  \citep{Thomas2005, deLaRosa2011}.
Thus, a high-$[\alpha/Fe]$ value would result from a concentrated burst of star formation.
Consequently, in the high-$[\alpha/Fe]$ environment, a relatively younger SN Ia progenitor will form. 
\citet{Nicolas2021} and \citet{Ginolin2025} showed that when an SN Ia explodes in younger environments, which is more likely to have a younger progenitor star, the value of $x_1$ is higher than in older environments. 
Considering the correlation between $x_1$ and \nickel{} mass (Eq.~\ref{eq:ni_x1}), this progenitor in the high-$[\alpha/Fe]$ environment will likely produce more \nickel{} mass. 
However, there are no studies about the impact of $(\alpha/Fe)_{progenitor}$ on the SN Ia luminosity, while there are some theoretical studies to investigate the impact of the progenitor star's metallicity on the SN Ia luminosity (e.g., \citetalias{Timmes2003}, \citealt{Kasen2009}). 
Hence, it would be an interesting test to explore the effect of $(\alpha/Fe)_{progenitor}$ on the SN Ia explosion.

Our finding in the relation between \prometal{} and HR is inconsistent with previous studies using host galaxy metallicities.
Previous studies mainly used the currently-observed status of host galaxies, which might be different from the progenitor star's environments.
In addition, previous studies typically used the gas-phase metallicity $(O/H)$ inferred from the stellar mass of host galaxies, not \prometal{}.
Also, it is known that the light-curve standardization process over-corrects the brightness of SNe Ia.
Therefore, our result using \prometal{} will be more robust.
However, we need more data to confirm our findings.

We expect more data to come from ongoing and upcoming surveys, such as the Zwicky Transient Facility \citep[ZTF;][]{Bellm2019, Graham2019} and the Rubin Observatory’s Legacy Survey of Space and Time \citep{lsst2009}. 
ZTF SN Ia DR2 released 3628 spectroscopically classified SNe Ia at $z < 0.3$ \citep{Rigault2025}.
2663 SNe Ia (out of 3628) passed the basic cuts for cosmological measurements.
We expect that half of them will explode in early-type or passive host galaxies \citep[see][for the rate of SNe Ia at $z<0.1$]{Rigault2020}.
This means we will have $O(10^{3})$ SNe Ia and host galaxies.
This number will be enough to obtain statistically significant results.

A holistic approach (from the SN Ia progenitor star to the explosion with host galaxy analysis and the SN Ia observation) to understand the scatter in the SN Ia luminosity and, thus, the underlying physics of SNe Ia is required to use SNe Ia as more accurate and precise standard candles.
This holistic understanding of the luminosity variation of SNe Ia in different environments or from different populations will provide an astrophysical solution to the Hubble tension.
For example, considering the SNe Ia in different environments (e.g., locally star-forming versus locally passive), which show a difference of $0.094\pm0.025$ mag in their corrected luminosity, \citet{Rigault2015} found a value of $H_{0}$ = $70.6 \pm 2.6$ $km$ $s^{-1}$ $Mpc^{-1}$.
This value shows no tension with the Planck measurement of $H_{0}$ = $67.4 \pm 0.5$ $km$ $s^{-1}$ $Mpc^{-1}$ \citep{Planck2020}.

\begin{acknowledgements}
We would like to thank the referee for the careful reading of the manuscript and for many constructive suggestions, which significantly expanded the manuscript.

Y.-L.K. was supported by the Lee Wonchul Fellowship, funded through the BK21 Fostering Outstanding Universities for Research (FOUR) Program (grant No. 4120200513819) and the National Research Foundation of Korea to the Center for Galaxy Evolution Research (RS-2022-NR070872, RS-2022-NR070525).

C.C. was supported by Basic Science Research Program through the National Research Foundation of Korea (NRF) funded by the Ministry of Education (RS-2022-NR070872) and by the National Research Foundation of Korea (NRF) grant funded by the Korea government (MSIT) (RS-2022-NR070525).

Y.C.K. was supported by Basic Science Research Program through the National Research Foundation of Korea (NRF) funded by the Ministry of Education, Science and Technology (NRF-2017R1D1A1B05028009).

This work used \textsc{\texttt{pandas}}\citep{McKinney2010}, \textsc{\texttt{numpy}}\citep{Harris2020}, and \textsc{\texttt{matplotlib}}\citep{Hunter2007}.
We also use \texttt{uncertainties}: a Python package for calculations with uncertainties, Eric O. LEBIGOT, \href{http://pythonhosted.org/uncertainties/}{http://pythonhosted.org/uncertainties/} and the LINMIX package \href{https://github.com/jmeyers314/linmix/}{https://github.com/jmeyers314/linmix/}.

\end{acknowledgements}

\begin{appendix}

\section{Correlation between \prometal{} and $(Fe/H)_{progenitor}$ and $(\alpha/Fe)_{progenitor}$}
\label{app:proz_feh_afe}

\begin{figure}
	\centering	
	\includegraphics[width=0.7\columnwidth]{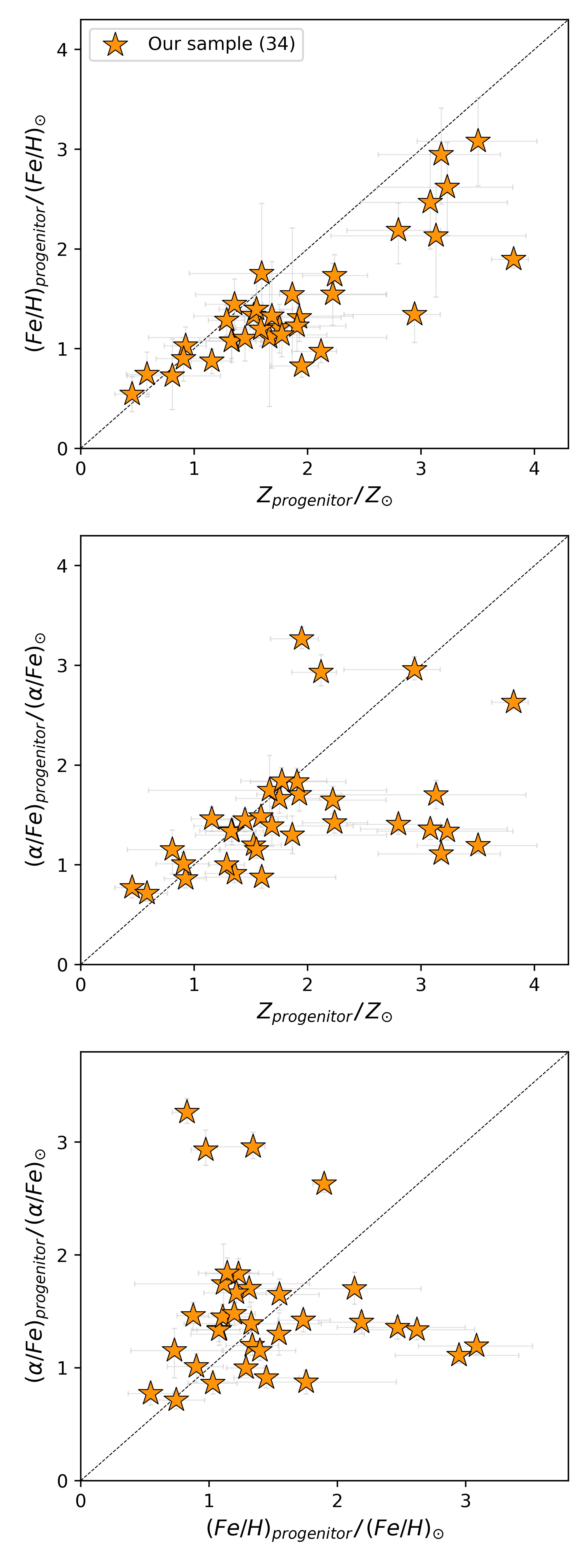}
	\caption{
	Correlation between \prometal{}, $(Fe/H)_{progenitor}$ and $(\alpha/Fe)_{progenitor}$ for our sample.
	Dashed lines indicate a one-to-one relation.
	} 
	\label{fig:proz_feh_afe}
\end{figure}

In the present work, we derived \prometal{} from $(Fe/H)_{progenitor}$ and $(\alpha/Fe)_{progenitor}$ via Eq.~\ref{eq:tbt03}.
Here, to see this visually, we plot the correlation between \prometal{}, $(Fe/H)_{progenitor}$ and $(\alpha/Fe)_{progenitor}$ for our sample in Fig.~\ref{fig:proz_feh_afe}.
Up to 2 \solarmetal{}, they appear to be a one-to-one relationship, but in the higher metallicity region, they show a large scatter.
In the figure, each point represents not only the SN Ia progenitor star but also the birth environment of the progenitor star.
Moreover, by the definition of \citetalias{Kim2024}, each point also indicates a whole galaxy at 0 Gyr old.
A detailed analysis and interpretation of this trend is beyond the scope of the present work, but would be of interest to the extragalactic community.

\end{appendix}

\end{document}